# Coupling phenomena and collective effects in resonant meta-molecules supporting plasmonic and magnetic functionalities: a review

Nicolò Maccaferri
*Physics and Materials Science Research Unit, University of Luxembourg, L-1511 Luxembourg, Luxembourg*
*\*nicolo.maccaferri@uni.lu*

**We review both the fundamental aspects and the applications of functional magneto-optic and opto-magnetic metamaterials displaying collective and coupling effects on the nanoscale, where the concepts of optics and magnetism merge to produce unconventional phenomena. The use of magnetic materials instead of the usual noble metals allows for an additional degree of freedom for the control of electromagnetic field properties, as well as it allows light to interact with the spins of the electrons and to actively manipulate the magnetic properties of such nanomaterials. In this context, we explore the concepts of near-field coupling of plasmon modes in magnetic meta-molecules, as well as the effect of excitation of surface lattice resonances in magneto-plasmonic crystals. Moreover, we discuss how these coupling effects can be exploited to artificially enhance optical magnetism in plasmonic meta-molecules and crystals. Finally, we highlight some of the present challenges and provide a perspective on future directions of the research towards photon-driven fast and efficient nanotechnologies bridging magnetism and optics beyond current limits.**

## Introduction

Plasmonics exploits the collective motion of conduction electrons in metals (plasmons), thus enabling light to couple with nanoscale objects, with the consequent generation of a plenty of novel and unexpected optical effects and functionalities. Plasmonic nanomaterials have been deeply studied in the last decade due to their crucial impact on several areas of nanoscience and nanotechnology [1]. Their unrivalled capability to squeeze light well beyond its diffraction limit, leading to extremely confined and enhanced electromagnetic fields on the nanoscale at optical frequencies [2- 7], is of great interest for the prospect of real-life applications [8], such as energy harvesting and photovoltaics [9- 12], wave-guiding and lasing [13], optoelectronics [14], fluorescence emission enhancement [15], plasmon-assisted bio-interfaces [16- 18] and nanomedicine [19- 21]. In this framework, metallic materials have been molded with the most cutting-edge nanofabrication techniques in order to create electromagnetically coupled nanostructured scatterers (that is, resonators, such as optical antennas [ 22 ]), dubbed metamaterials [23- 25]. By using the controllable amplitude, phase or polarization changes associated with the optical excitation of plasmons in structured meta-surfaces and plasmonic crystals, it is possible to reach an unprecedented and arbitrary manipulation of the optical wave-fronts in the near-field with subwavelength resolution and a strong modification of the light transmitted and reflected in the far-field [26,27]. Multifunctional metamaterials enabling a dynamic manipulation of light properties on the nanoscale are key features for future photonic nanodevices, which use light as an information carrier in optical communications, sensing, and imaging [28]. In this scenario, meta-surfaces' building blocks, named meta-atoms or meta-molecules, which combine magnetic and plasmonic functionalities, represent a promising route for enabling active nanophotonics and flat-optics [29]. For a long time, it has been deemed impossible to combine light and magnetism because of a frequency gap where light moves 10,000 times faster than magnetism reacts, which means they do not feel each other and cannot integrate. By capturing the light using metamaterials, it is possible for the two to interact on the nanoscale. In this artificially created meta-surfaces plasmons can be connected with magnetic materials via different types of magneto-optical effects.

The aim of this Review is to describe the main physical phenomena taking places when meta-atoms supporting localized surface plasmons interact each other through near-field interactions (meta-molecules) or far-field diffraction coupling (meta-crystals), and contextualize recent works on such metamaterials displaying either magneto-optical functionalities coming from the magnetic nature of the constituent meta-atoms or artificial optical magnetism coming from their nanoscale patterning that mimics current loops, thus inducing oscillating magnetic fields at optical frequencies. In this context, we also discuss the impact of opto-magnetism on achieving an active control of magneto-plasmonic metamaterials properties through a dynamic manipulation of their inner electronic structure. The harnessing of all functionalities presented in this review might be the start point towards the development of a totally new class of dynamic metamaterials, which will eventually let light becoming arbitrary steerable using magnetism, and vice versa, thus eliminate the frequency gap. We also consider different applications of all the phenomena reported here, such as extremely sensitive bio- and chemical sensing, and future light-driven meta-magnetic memories.

## Localized plasmons and magneto-optics in magnetic meta-atoms

Plasmon modes can be subdivided into two classes: propagating surface plasmons (PSPs) and localized surface plasmons (LSPs). In this Review, we do not treat PSPs, also because they have been extensively studied, see for instance refs [2,30]. Nevertheless, it is worth mentioning that there is an active research community studying PSPs in magneto-plasmonics [31,32], whose applications span from the refractive index sensing and the detection and study of biomolecular binding events involving target analytes such as proteins, DNA, drugs, etc., [33-38] to metrology [39,40] or modulated color displays [41].

However, here we want to give to the reader a complete description of LSPs in magnetic meta-atoms, since they are the essential building blocks of either magneto-plasmonic crystals (MPCs) or meta-molecules (hetero-dimers/trimers or, more in general, heteromers), which are the systems displaying the collective effects and coupling phenomena treated in this work. In practical experiments, LSPs result from the coupling of metallic nanoparticles with, for instance, free-space radiation. They are, essentially, a combination of the oscillation of the free electrons in plasmonic nanostructures and the associated oscillations of the re-emitted electromagnetic field. From a mathematical point of view, they can be described classically as damped harmonic oscillators, whose resonance frequency depends on the size, shape, composition, and local optical environment of the particle [42,43]. LSPs resonances (LSPRs) typically occur in the visible to near-infrared part of the spectrum for nanostructures of noble metals (Au, Ag, Cu). Aluminum can also be used as viable plasmonic material supporting resonances in the UV–visible region. [44-46]. More recently, heavily doped semiconductors [47] allowed to expand the research in the mid-IR spectral range [48-50], which is the fingerprint region of molecular vibrations. In this framework, and thank to the development of simple fabrication routes for the metal nanoparticles (e.g., by colloidal chemistry based on reduction of metal salts [51]) and nanoparticle arrays (e.g., by nanosphere colloidal lithography [52-56]), LSPs have been found to be suitable for a wide range of applications, including subwavelength imaging and superlensing [57-61], nanolasing [62-64], light trapping and concentrators [65-67], plasmon-enhanced optical tweezers [68-70] supersensitive plasmonic metamaterials sensors [71-74], improved photovoltaic devices [75] active optical elements [76,77] etc. In this framework, magnetism have emerged as a valuable route to add an extra degree of freedom to plasmonics, since it allows to actively induce significant changes in the optical response of meta-atoms either entirely [78- 85]or partially [86- 91] made of magnetic materials and supporting LSPs (for a detailed overview we refer the reader to the reviews by Maksymov [92], and Pineider and Sangregorio [93]). More in detail, magnetic materials possess what is called a magneto-optical (MO) activity, arising from spin-orbit coupling of electrons, resulting in a weak magnetic-field induced modulation of the intensity and/or polarization of the reflected and transmitted light. The unique optical properties of magneto-plasmonic meta-surfaces composed by magnetic meta-atoms, which behave as nanoantennas, arise from combining strong local enhancements of electromagnetic fields via surface plasmon excitations with their inherent MO activity. For a circular disk-like magneto-plasmonic meta-atom, incident radiation of proper wavelength excites a LPSR. When the nanoantenna is *magnetic-field (**H**) activated*, a second LPR is induced by the inherent MO activity. The MO-induced LSPR (MOLSPR) is driven by the optical LSPR in a direction orthogonal to both ***H*** and the LSPR itself. The control of the relative amplitude and phase lag between these two orthogonal resonant electric dipoles determines a magnetic-field induced active control of the polarization of the interacting light [ 94 - 95 ]. In the past decade, magneto-plasmonic nanoantennas-based materials and crystals were intensively investigated as promising systems for the realization of 2D flat-optics non-reciprocal nanodevices, such as rotators, modulators, and isolators [96- 101] and to enhance and manipulate magnetic-induced optical activity on the nanoscale [102- 110]. Furthermore, magneto-plasmonic meta-atoms were proven to provide an unprecedented accuracy in refractive index sensing [111] and label-free detection of single-molecules [112], as well as in single cell mechanogenetics [113], bioimaging and sensing [114] and nano-medicine [115,116]. Finally, magneto-plasmonic meta-atoms have been recently found to have an interesting impact also on thermo-plasmonic applications [117] or in the integration with 2D materials [118]. It is worth mentioning here that with single magneto-plasmonic meta-atoms remarkable enhancements up to 1-order of magnitude compared to what achievable without the assistance of plasmons, as in the corresponding continuous film, have been reported. However, this already notable enhancement of the magneto-optical response is still too low for practical uses, and this has hindered the utilization of magneto-plasmonics to active nanophotonics. In addition, the quality factor Q of LSPRs, which is intrinsically linked to the ratio of energy stored to the energy lost by the oscillator and can be estimated as $Q = \lambda_0/\Delta\lambda$ ($\lambda_0$ is the resonance wavelength, and $\Delta\lambda$ is the width of the resonance), is found to be of the order of 10 for noble-metal plasmonic nanostructures, a value well below that desired for many applications. Being magneto-plasmonic meta-atoms composed by magnetic materials such as Ni, Co or Fe, the losses are even larger, and so Q is even smaller than 10 in this case. Furthermore, experimentally a typical LSPR is broader than the aforementioned simple formula for Q predicts, owing to radiative damping and dynamic depolarization (the effect of retardation within the particle). Finally, being Q, to a first order approximation, independent of the geometric form of the nanostructure and of the dielectric medium that surrounds the nanostructure itself, and dependent only on the dielectric function of the metal at the plasmon frequency [119], several strategies have been developed so far to overcome these limitations. Fortunately, and at first sight rather surprisingly, the restrictions placed by both the low Q-factor and the limited enhancement of the magneto-optical effects in individual (randomly arranged) meta-atoms, can be largely overcome when magneto-plasmonic meta-atoms are arranged in either heteromeric structures or arrays. In these configurations, the electromagnetic fields related to the excitation of a LSPR of one meta-atoms can influence the response of the neighboring meta-atoms. This electromagnetic coupling can take several forms: via either near- or far-field coupling.

## Near-field interactions in magneto-active meta-molecules

Meta-atoms interact via near-field coupling when they are relatively closely packed, that is the inter-particle gap distance is of the order of 10nm, leading to significant spectral shifts of the LSPRs and a modification of their line-shapes due to the hybridization of the modes [120,121]. The simplest of these systems, namely a plasmonic dimer [122,123], can found several applications, such as in near-field mediated strong-coupling interactions [124-126], ultrasensitive bio-detection [127] or to study biological processes inside cells through enhanced spectroscopy techniques [128]. In the framework of magneto-plasmonics, such near-field interactions were used to increase the control on a plenty of phenomena taking place in magneto-plasmonic meta-molecules. Armelles et al. [129] showed that the interaction between a gold and magnetic nano-disks organized in a vertical dimer fashion with a dielectric spacer leads to the appearance of magneto-optical activity in the gold disk induced by the magnetic one. Moreover, at specific wavelengths the interaction between these two nano-disks cancels the net electromagnetic field in the whole meta-molecule, strongly inhibiting the magneto-optical activity of the whole system, giving rise to a magneto-optical-induced of electromagnetic transparency (see Fig. 1). The same group demonstrated also that De Sousa et al. showed that by proper design of the dipole-dipole interaction in this meta-molecule it is possible to generate an induced a magneto-optical activity in the gold meta-atom even larger than that of the magnetic one [130]. On the same line, Zubritskaya et al. [131] demonstrated that magneto-plasmonic dimers made of pure nickel can sense nanoscale distances with an extremely high sensitivity and proposed a concept to optically measure the nanoscale response to the controlled application of force with a magnetic field. Recently, some of the authors of the aforementioned work demonstrated that meta-molecules composed by one nickel meta-atom closelyenable an unprecedented magnetic field-induced modulation of the far-field chiroptical response with this surface exceeding 100% in the visible and near-infrared spectral ranges, opening the route for nanometer-thin magneto-plasmonic light-modulating surfaces tuned in real time and featuring a broad spectral response [132] (see Fig. 1). Similarly, to enhance the magneto-optical response, Sachan et al. [133] looked at a novel way to boost the strength of the response of individual magnetic nanoparticles by combining two materials, one metallic and the other ferromagnetic, to make individual nanostructures supporting the so-called ferroplasmons [134]. At the peak positions of the hybridized LSPRs associated with these heterodimers, the near-field enhancement generated in the gap between the meta-atoms was one order of magnitude bigger than the enhancement produced using isolated metallic spheres. A similar approach to enhanced the optical and magneto-optical response of nanoferromagnets was recently proposed by López-Ortega et al. [135], where they showed that magneto-plasmonic core/shell nanocrystals, where the noble-metal and magnetic components are placed in the core and the shell regions, respectively, can enable at least a 5-fold enhancement of the magneto-optical response compared to dispersed pure ferromagnetic nano-particles (see Fig. 2). A similar system can be also used to transfer magnetic information inside the noble-metal structure, as demonstrated by Pineider et al. [136], who reported the direct observation of spin-polarization transfer across colloidal magneto-plasmonic Au/Fe-oxide core/shell meta-molecules relying on direct hybridization between the Au and Fe states at the core/shell interface. It would be indeed interesting, although really challenging, to understand if this type of mechanism can be photo-activated and then controlled with either light and/or other external agents. Some promising results in this direction were recently reported by Yin et al. [137], who showed that plasmon–exciton coupling in degenerately doped $In_2O_3$ nanocrystals allows for the indirect excitation of magnetoplasmonic modes with a subsequent splitting of the excitonic states (see Fig. 2). Moreover, recently Yu et al. [138] have studied Fano switching in coated magnetoplasmonic nanoparticle with a Kerr-type non-linear plasmonic shell and a magnetic core featuring two plasmon dipolar modes, which hybridize in a nonlinear regime exhibiting optical tri-stable response. These results pave the way for enabling a dynamic and selective light-driven control of electronic spin polarization for future photo-reactive spintronic devices [139].

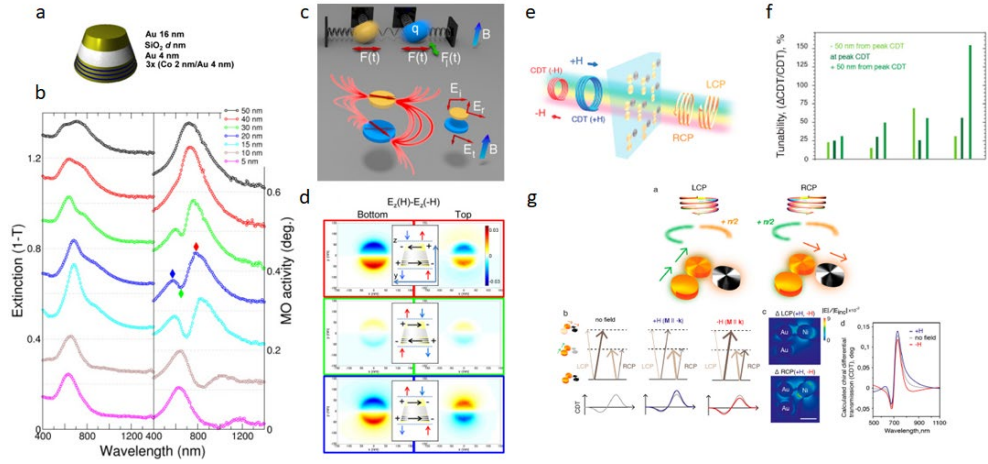

Fig 1. (a) Schematic drawing of the nanoresonators composed of a purely plasmonic Au disk and a magnetoplasmonic Au/Co superlattice disk separated by a dielectric spacer. (b) Extinction (left) and MO activity (right) spectra of the nano-resonators as a function of $SiO_2$ thickness. The different dashed horizontal lines indicate the zero value for each spectrum above the line. The blue, green and red diamonds appearing in the experimental MO response for 20 nm $SiO_2$ correspond to the spectral positions where the $E_z$ field distributions are calculated. (c) Top side: spring model representing a two coupled masses system excited by an harmonic force, F(t), along x axis: one of the masses (blue) is charged (q) and a static magnetic field (B) is applied along the z direction, inducing a Lorentz force, $F_l(t)$, along the y direction. The y-movement is transferred to the other mass through the coupling. Bottom side: two interacting electric dipoles, representing two metallic disks, excited by an incident beam polarized along the x axis: one of the disks (blue) has magneto-optical activity and a static magnetic field (B) applied along the z direction induces a rotation of its electric dipole, which is transferred to the other dipole through the interaction. The rotation modifies the polarization direction of the reflected and transmitted light. (d) Calculated near field intensity of the $E_z$ component for a nanoresonator with 20 nm $SiO_2$ spacer in two planes above the top disk and below the bottom one, with the incident field polarized along the x direction in presence of an external magnetic field. The red (blue) arrows in the insets represent the positive (negative) values of the Ez field component. Difference of the $E_z$ components for magnetic saturation along opposite directions in the same planes and for the same structure. This difference accounts for the effect of the applied magnetic field: the appearance of a dipole along the y direction for both the magnetoplasmonic (intrinsic dipole) and the plasmonic (induced dipole) disks. Adapted from ref. [129]. (e) Magnetic tunability of the chiroptical transmission. (a) Schematics of magnetically controlled chiral differential transmission (CDT) through a surface of magneto-plasmonic trimer nano-antennas. (f) Magnetic tunability of the surfaces studied in ref. [132], expressed as (ΔCDT/CDT) at different wavelengths: the peak CDT wavelength (central bar), and the wavelengths +50 nm (right bar) and −50 nm from the peak CDT wavelengths (left bar). (g) The principles of chiroptics in symmetric bimetallic nanoantennas. Top panel: chiroptical response of a bimetallic trimer antenna with highlighted electromagnetic pairs and their longitudinal resonances (green and orange), excited with quarter-period phase delay. Bottom-left panel: chiroptical response (and magneto-chiral response of a bimetallic trimer antenna with magnetization direction antiparallel (middle panel) and parallel (right panel) to the light propagation direction (note the schematics of the nanoantennas resonances to the left). Bottom-right panel: FDTD-calculated magnetically modulated electric near-field of bimetallic Au−Au−Ni trimer antenna at 690 nm for LCP and RCP light (left, scale bar 100 nm) and calculated magnetically modulated CDT of the bimetallic trimer antenna (right). Adapted from ref. [132].

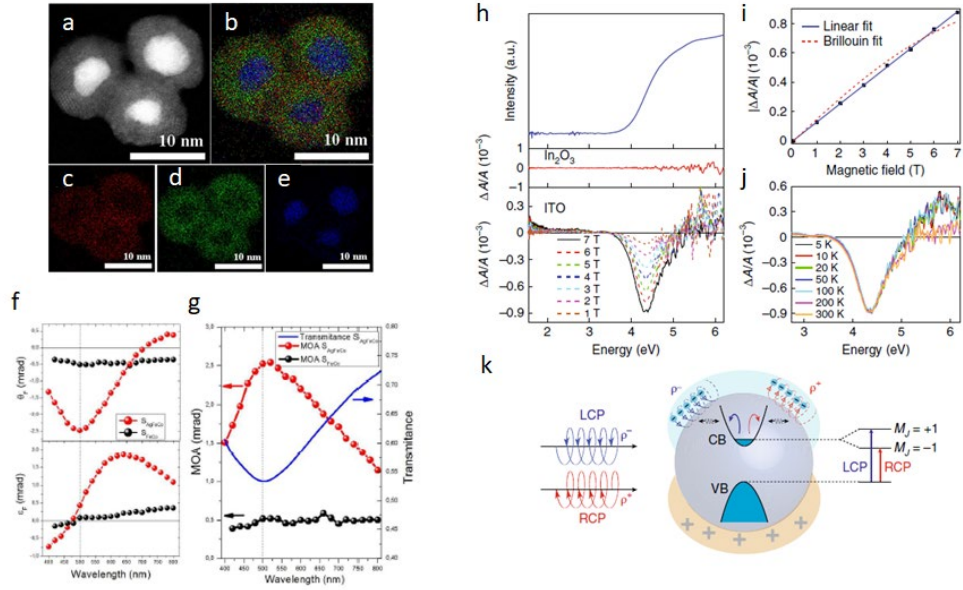

Fig. 2. (a) High resolution STEM-HAADF image of sample SAgFeCo and (b) the respectively overlay EDS image for the (c) Fe K edge, (d) Co K edge and (e) Ag L edge. (f) Faraday rotation (top) and ellipticity (bottom), and (g) MO activity spectra for samples SAgFeCo and SFeCo. Blue curve in panel (g) shows the transmission spectra of sample SAgFeCo for comparison. Adapted from ref. [135]. (h) Absorption (solid blue line) and MCD (solid black and coloured dashed lines) spectra of ITO nanocrystals (that contain 10% Sn4+) collected at 300 K. MCD spectra were recorded at different external magnetic field strengths, as indicated. The 300 K MCD spectrum of $In_2O_3$ nanocrystals (collected at 7 T) is shown for comparison (solid red line). (i) Magnetic field dependence of MCD intensity at 4.34 eV for ITO nanocrystals in a as a function of the magnetic field strength. (j) The 7 T MCD spectra of nanocrystals in a collected at different temperatures. (k) Schematic representation of the splitting of the conduction band (CB) states induced by angular momentum of the cyclotron magnetoplasmonic modes. On excitation with LCP and RCP light in a magnetic field, cyclotron magnetoplasmonic modes with helicities ρ − (curved dashed blue line) and ρ + (curved dashed red line), respectively, are formed. These modes couple with the exciton and transfer angular momentum (blue and red curved arrows) to the CB excited states, which causes their splitting ($M_J = \pm 1$) and the difference in absorption of LCP (vertical blue arrow) and RCP (vertical red arrow) light. Valence band (VB) states are not subject to splitting. Adapted from ref. [137].

## Far-field diffractive coupling in magneto-plasmonic crystals

Beyond isolated or incoherently interacting meta-molecules made of near-field coupled homo- or hetero-meric building-blocks, meta-atoms/molecules can be orderly arranged in lattices whose period is comparable to the wavelength of the incident light. In such a configuration they scatter light to produce diffracted waves in the plane of the array which couple to the LSPRs associated with individual meta-atoms or unit cell of meta-molecules. By using the right combination of meta-atoms' size and shape or meta-molecule arrangement, together with an appropriate array period, one can design a system where the light scattered by each unit cell of the lattice is in phase with the LSPR induced in its neighbor, thereby acting to counter the damping of the response of the single lattice unit, thus reinforcing the resonance in the neighboring particle. This approach has been found to increase significantly the Q of the resonance, lying on a remarkably narrow (down to a few nm) resonances called surface lattice resonances (SLRs), thus representing a dramatic improvement in terms of FWHM, which can be shrink of at least one order of magnitude compared to typical isolated meta-atoms' resonances. Furthermore, these SLRs lead to a dramatic enhancement of both absorption and the local electric fields near the meta-atoms/molecules; all of these phenomena are important for a variety of projected applications, also in magneto-plasmonics, to enhance even more the magneto-optical response of magneto-active meta-atoms. For a detailed review of the basic physical principles and properties of plasmonic SLRs and their applications, we refer the reader to the recent work by Kravets et al. [140]. In the present review, we otherwise want to highlight the potential that SLRs offer to enhance magnetic effects by directly using magnetic or hybrid noble-metal/magnetic materials. One of the first works reporting on arrays of magneto-plasmonic meta-atoms is undeniably that by Du et al. [141], who presented systematic studies of the magneto-optical activity of Au/[Co/Pt]xN/Au (N is an integer number related to the number of bi-layers Co/Pt) nano-disk arrays with various disk sizes, shapes and lattice constants. Being the rectangular square lattice studied in ref. [141] composed of noble-metal/ferromagnetic composite particles, the effects arising from the pure plasmonic response of the magnetic material was masked. Furthermore, the array period was chosen such that sharp SLRs observed in other works were not present (see Fig. 3). An interesting study was later performed by Kataja et al. [142], who studied magneto-plasmonic crystals made of pure nickel meta-atoms. When breaking the symmetry of the lattice, they find that while the frequency of the SLRs in the optical response is determined by the periodicity orthogonal to the polarization of the incident field, in striking contrast, the magneto-optical Kerr response is controlled by the period in the parallel direction. They showed that spectral separation of the response for longitudinal and orthogonal excitations provides versatile tuning of narrow and intense magneto-optical resonances, which were enhanced 3-fold when compared with a system of randomly oriented particles. The same authors demonstrated also that SLRs significantly enhance magnetic-induced circular dichroism signal compared to randomly distributed nickel meta-atoms [143]. Similarly, arrays of elliptical nickel nanoantennas were studied by Maccaferri et al. [144], who showed that the diffractive coupling in these arrays is dictated by two orthogonal and spectrally detuned in-plane LSPRs of the individual building blocks, one directly induced by the incident light, the other produced through the application of an external magnetic field. This approach leads to highly tunable and enhancement of almost one order of magnitude of the magneto-optical response when compared with a continuous Ni film or surfaces made from randomly distributed magneto-plasmonic elliptical nano-antennas (see Fig. 3). In another study, Kataja et al. [145] investigated optical and the magneto-optical response of Au and Ni nanoparticles organized in a check-board configuration. Analysis of the optical fields indicated that both the Au and Ni nanoparticles contribute to SLRs and thus to the magneto-optical activity of these hybrid arrays. In addition, it was shown that both the magneto-optical and the optical response of the hybrid arrays could be adjusted by altering the size of the gold nanoparticles. Similarly to the results reported in refs. [129,130], also in this configuration the magnetic field-induced SLRs in the nickel nano-disks promote an enhancement of those in the gold ones, thus demonstrating that this approach may lead to new possibilities for the realization of tunable hybrid magneto-plasmonic crystals. A similar approach was also proposed previously by Caminale et al. [146], who investigated the magneto-optical response of chemically synthesized iron oxide magnetic nanocrystals, optically coupled with ordered planar arrays of plasmonic meta-atoms, where they were able to arbitrary superimpose or detune the LSPRs of the nanoparticles arrays with respect to the dominant magneto-optical resonances of the magnetic meta-atoms to either merge or separate the purely plasmonic and the magnetic contributions in the magneto-optical spectrum to tune the enhancement of the magneto-optical signal. Recently, Pourjamal et al. [147] proposed to arrange meta-molecules in an ordered fashion. They demonstrated that the magneto-optical response of square arrays of vertical dimers composed by Ni, SiO$_2$ (as spacer) and Au, similar to the randomly distributed dimers reported in ref. [129], is governed by a complex interplay of near- and far-field interactions.

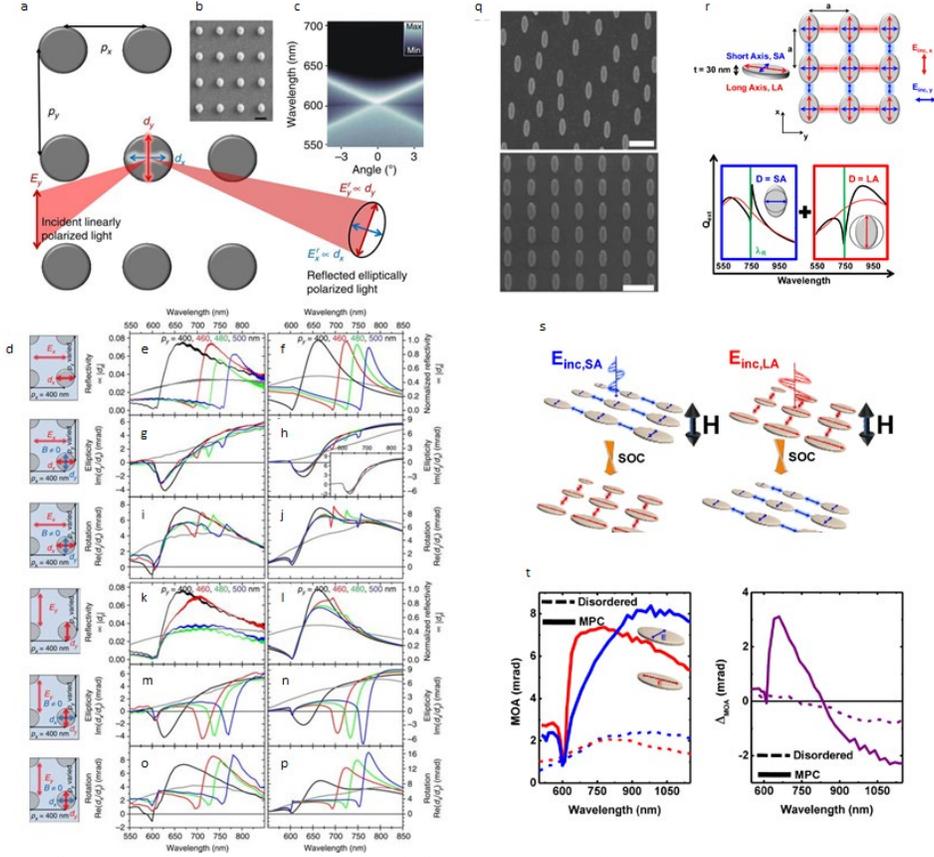

Fig. 3. (a) Schematic of the system studied in ref. [142]. In the presence of magnetic material, the system response is governed not only by the induced dipole moment $d_y$ parallel to the driving field $E_y$ and the lattice period $p_x$ (direction of dipole radiation), but also by the spin–orbit-induced and magnetic-field tunable dipole moment $d_x$ and lattice period $p_y$. (b) Scanning electron micrograph of an ordered rectangular array of cylindrical Ni nanoparticles. Scale bar, 200 nm. (c) Angle- and wavelength-resolved optical transmission of a sample with $p_x=p_y=400$ nm and with particle diameter 120 nm, showing crossing of the <+1, 0> and <−1, 0> diffracted orders of the lattice at normal incidence. (d) Schematics illustrating the direction of incident polarization and the induced dipole moments that are probed in the measurements. (e,f) Normal incidence optical reflectivity for incident light polarization $E_x$. The magneto-optical Kerr ellipticity (g,h) and rotation (i,j) with polarization $E_x$. Normal incidence optical reflectivity (k,l), the magneto-optical Kerr ellipticity (m,n) and rotation (o,p) for polarization $E_y$. The graphs show experimental data (left column) and results from DDA simulations (right column). The black, red, green and blue curves correspond to the periodicities $p_y=400, 460, 480$ and 500 nm, in all the figures. The grey line corresponds to a random sample. Adapted from ref. [142]. (q) SEM image of a portion of the elliptical nanoantennas on glass randomly distributed (top) and in the MPC (bottom). Short axis (SA) = 100 nm, long axis (LA) = 250 nm, and thickness t = 30 nm. The pitch of the MPC is 400 nm in both LA and SA directions and the filling factor of both samples is 12%; scale bars 500 nm. (r) schematics of the system studied in this work, where the single nickel nanoparticles displays two different optical responses along the two in-plane principal x- and y-axis, aligned with the nanodisk LA = 180 nm and SA = 100 nm, respectively. The thickness of the single antenna is 30 nm. (b) Right panel: extinction efficiency $Q_{ext}$ spectra of the magneto-plasmonic crystal sketched in the top panel (black curves). The red curves indicate the LSPR of an isolated nano-disk. The vertical green solid lines indicate the spectral position of the Rayleigh's anomaly $\lambda_R$. The system is assumed to be immersed in a homogeneous refractive index n = 1.5. All the curves were calculated using the Coupled Dipole Approximation (CDA). (s) sketch of the same system shown in (r) when an external static magnetic field is applied along the direction perpendicular to the lattice plane and for incident field $E_{inc}$ applied along the SA (blue incident radiation) and LA (red incident radiation). The spin-orbit coupling activated by the magnetic field induces dipolar oscillations perpendicular to the direction of the incident illumination (MO-LSPRs), leading to a predominant far-field interaction parallel to the polarization of $E_{inc}$ (MO-SLRs). (t) Experimental MO activity spectra measured for the system of randomly arranged elliptical antennas (dashed lines) and for the magneto-plasmonic crystal (solid lines) for both SA-(blue lines) and LA-(red lines) illumination. Magneto-optical anisotropy ΔMOA for the random (violet dashed line) and MPC (violet solid line) cases. Adapted from ref. [144].

In this case the magnetic field-induced LSPR induced in the gold building block was found to be higher than its magnetic counterpart, leading to a higher MO response compared to that of the check-board configuration of ref [145]. Similarly, recently a hybrid system made of ordered noble metal/ferromagnets were studied also by Atmatzakis et al. [148], who demonstrate enhanced resonant Faraday polarization rotation in plasmonic arrays of bimetallic nano-ring resonators consisting of Au and Ni sections (see also Fig. 4). The design proposed allows the optimization of the trade-off between the enhancement of magneto-optical effects and plasmonic dissipation, and, although nickel filling factor was ~6% of the total surface of the metamaterial, the magneto-optically induced polarization rotation is equal to that of a continuous nickel film. Finally, Chen et al. [149,150] have recently reported the interesting possibility to couple SLRs and PSPs in the same system by using a pure ferromagnetic magneto-plasmonic crystal consisting of two-dimensional ordered Ni nano-disks array on Co film (see Fig. 4).

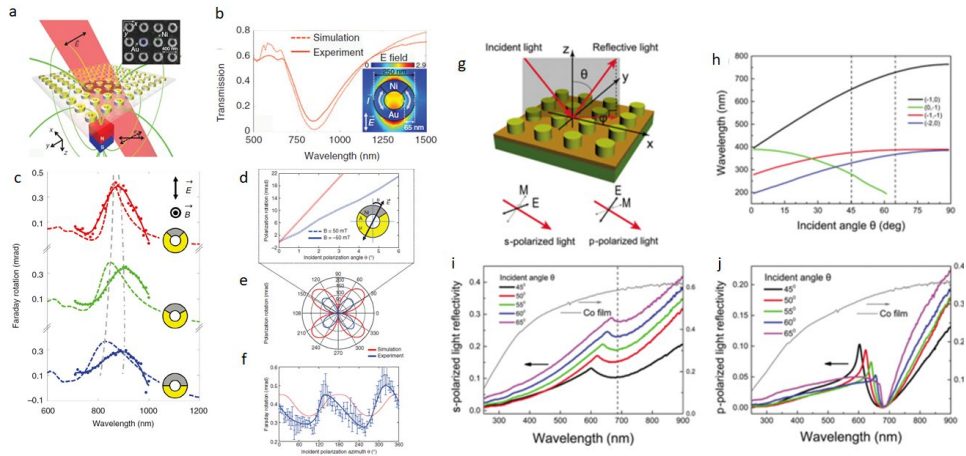

Fig. 4. (a) Schematic of a bimetallic metamaterial array. In the presence of an external magnetic field (represented by the green lines), the meta-material induces rotation of the polarization azimuth angle, ϕ, on the incident beam. The inset is a SEM image of a fabricated sample. (b) Characteristic transmission spectra of a metamaterial sample with Ni sector that spans 90° illuminated with light polarized in the symmetry plane of the ring. The dashed line is obtained by a computational analysis of an infinite 2D array, while the solid line corresponds to the experimentally measured spectra of a 100×100 μm$^2$ metamaterial sample. Inset: Electric field distribution in the vicinity of the bimetallic ring at resonance (λ=850 nm). (c) Faraday rotation of polarization azimuth. Dots describe the experimental results for samples with Ni sectors that span over 90° (red), 135° (green), and 180° (blue) in the presence of a 100-mT external magnetic field. Solid lines act as guide for the eye. Dashed lines represent the corresponding computational results when transmission is normalized to the experimental values. The dashed grey line follows the peak position of simulation data, while dot-dashed line shows the corresponding trend for the experimental results. (d,e) Rotation of the polarization azimuth of a beam transmitted through the sample (135° Ni sector). Red lines represent computational analysis, and blue lines stand for experimental results. Solid and dashed lines represent opposite directions of external magnetic fields. θ is the angle of polarization azimuth in reference to the symmetry axis of the metamaterial array. (f) The effect due to the presence of the external magnetic field appears as the difference between the solid and dashed lines plotted in (d) and (e). The red line corresponds to numerical results, while blue circles represent experimental measurements. The blue line is a guide for the eye. Adapted from ref. [148]. (g) Realization of Fano resonance in magneto-plasmonic crystal where a rectangular coordinate system is built on the sample; the x axis and y axis are along the basic directions of the square lattice. The relative orientation between the incident light and the sample is defined by the incident angle θ and azimuthal angle φ. Two configurations of polarization are explored: s polarized with the electric field perpendicular to the incident plane, and p polarized where the electric field lies in the incident plane. (h) The θ dependence of the calculated spectral positions to excite PSP modes for several diffraction orders (φ=0∘). (i,j) Measured reflectivity of the sample (left axis) at different θ and Co continuous film at θ=45∘ (right axis) versus wavelength for s-polarized light (i), and p-polarized light (j), maintaining φ=0∘. Adapted from ref. [150].

Regarding possible applications of these magneto-plasmonic crystals, one of the most interesting is sensing. It has been proven that nanostructure design [151], phase-sensitive detection schemes [73,112,152] and ordering of plasmonic nanoparticles into a regular array

[153,154] can be used to enhance the sensing figure of merit (FoM), which is a parameter defining the sensor performance and is equal to the ratio between the sensitivity of the resonance (shift of the LSPR peak Δλ divided by the refractive index change Δn or by the thickness Δt of the molecules adsorbed by the meta-atom surface) and the FWHM of the resonance [155-158]. Pourjamal et al. [159] combined all these aspect in one system, demonstrating that the hybrid magneto-plasmonic crystal studied in ref. [147] can be used for high-resolved and enhanced sensing of refractive index changes, proving experimentally that, compared to random distributions of pure Ni nano-disks or Ni/SiO$_2$/Au dimers or periodic arrays of Ni nano-disks, this hybrid system displays one order of magnitude larger sensing FoM. In other cases, the concept of FoM can be generalized in terms of the readout performance of detection of minimal effects, which is actually strictly related to the signal-to-noise ratio [160]. Related to the latter definition, Kikuchi and Tanaka performed a systematic study on the longitudinal Kerr effect in arrays of Au/Co/Au rectangular meta-atoms, demonstrating that the addition of plasmonic Au layer to the magnetic active Co layer is effective in terms of enhancement of both the optical and magneto-optical responses compared to bare Co nano-patch array [161].

Beyond sensing, from the technological point of view, one could envision potential applications of such magneto-active meta-molecules and meta-crystals in the field of magnetic data storage. Current hard disk devices with capacities of several terabits of information have data storage cells with dimensions of around $10^4$ nm$^2$. The reduction of grain sizes of magnetic materials to minimize the bit cell size could increase the capacities of current HDDs; a problem, however, is that storage for materials with grains below a certain size becomes unstable, while their magnetization may undergo arbitrary changes due to thermal effects [162,163]. When the temperature of a magnetic material that cannot be written to at room temperature is increased towards its Curie temperature, the magnetic field needed to change magnetization (the coercive field) may be reduced below the available switching field [164]. Energy-assisted magnetic recording [165] has been proposed to overcome this problem. Using metal nanoparticles to confine and enhance electromagnetic fields providing the energy for heating the magnetic material would seem the most logical approach because it requires no direct mechanical contact with the medium, and can be relatively easily integrated in a write head [166,167]. The combination of plasmon-induced heating by light with ferromagnetic nanoislands enables one to overcome limitations imposed by the grain size. Studies in this direction were also reported by Kataja et al. [168], who reported on the manipulation of magnetization by femtosecond laser pulses in a periodic array of cylindrical nickel nanoparticles. They showed a strong correlation between the excitation of SLRs and changes in magnetization. More in detail, they proved that excitation of SLRs triggers either demagnetization in zero magnetic field or magnetic switching in a small perpendicular field and explained both magnetic effects in terms of plasmon-induced heating of the nickel meta-atoms to their Curie temperature. In this framework, one can also envisage a route for designing new types of magneto-plasmonic metamaterials based on the use of ultrafast plasmon-induced career manipulation approaches [169] to reach an active thermo-plasmonic control of magnetic states in either artificially spin-ices [170,171] or bi-stable nano-magnetic switches [172] to face fundamental challenges of current technology, such as power dissipation and limits of scaling.

## Artificial optical-magnetism in coupled plasmonic metamaterials

We mentioned at the beginning of this article that there is an alternative approach to artificially achieve magnetic effects by using nanostructured meta-molecules. The simplest structure to generate magnetic modes is the metal-insulator-metal (MIM) structure [173-175], where the excitation of gap plasmons induce strong magnetic resonances [176].

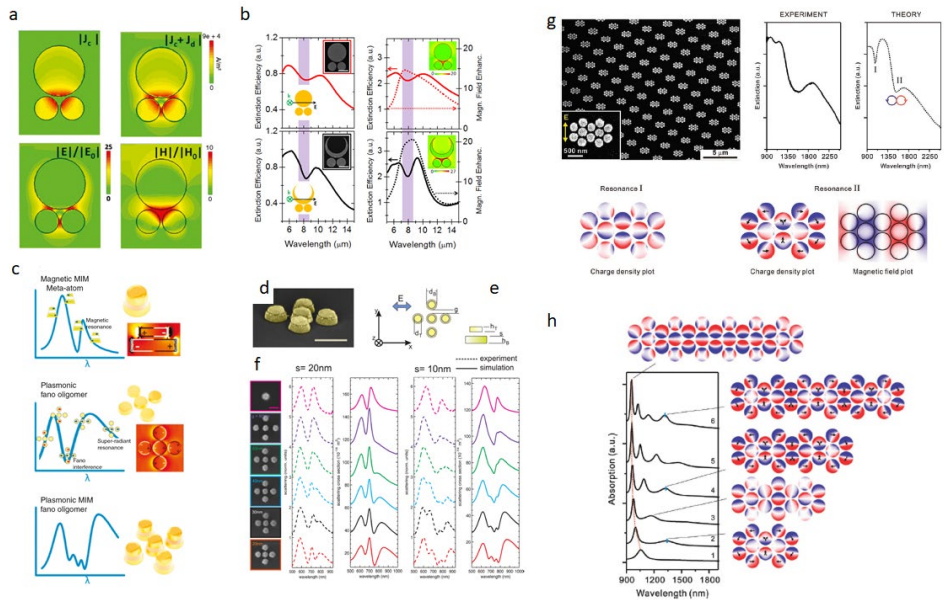

Fig. 5. (a) 2D plots of conduction current density (top left) and total charge current density, i.e. conduction plus displacement current (top right), together with the electric (bottom left) and magnetic field enhancement (bottom right) at the Fano resonance. Adapted from ref. [191]. Copyright 2014 American Chemical Society. (b) Left, experimental extinction efficiency spectra associated with the respective arrays of disk trimer (red curve) and MTR (black curve), for normal incidence condition (see the left side insets for the polarization direction). The insets respectively show the SEM images of a single disk trimer (upper inset) and a single moon trimer resonator (MTR) (lower inset) (scalebar: 1 μm). Right, respective simulated extinction efficiency spectra (solid lines) and magnetic field enhancement spectra (dotted lines) of the disk trimer array (red curves) and the MTR array (black curves). The insets show the 2D plots of the magnetic field enhancement distribution in the Fano resonance condition (λ=8 μm). Adapted from ref. [190]. (c) Scattering properties of a single vertical MIM dimer (top), plasmonic oligomer (centre) and MIM oligomer (bottom). The properties of the first two architectures are combined in the latter system that supports multiple subradiant modes, thus generating a quasi-broadband magnetic response at visible frequencies. (d) Tilted SEM image of an MIM pentamer (scale bar 200 nm). (e) A sketch of the simulated system with the nomenclature used in the text. Each dimer is constructed from two cylindrical gold nanodisks with similar aspect ratio (d/h). The incident plane wave enters from above and is polarized along x. (f) Representative experimental dark-field spectra (dashed lines) and simulated scattering spectra (continuous lines) of individual MIM pentamers for different in-plane gap distances g and for the two investigated spacer thicknesses, s=20 and 10 nm, together with single dimer spectra for comparison. The color-coding links the spectra to the SEM images shown in the left column. Adapted from ref. [185]. (g) Left: SEM image of a plasmonic naphthalene sample fabricated by electron-beam lithography. The periodicities in both directions are 2200 nm. The interparticle gap distance is 30 nm and the thickness of the gold particles is 80 nm. The diameter of the gold particles is 260 nm. Inset: enlarged view of the sample. Right: experimental and simulated extinction spectra of the plasmonic naphthalene sample. The light polarization is along the direction of the two shared gold particles. Bottom: charge density and magnetic field plots at resonances I and II. Resonance I corresponds to a double Fano resonance mode. Resonance II corresponds to a magnetic ring mode, in which antiphase magnetic plasmons are excited. (h) Simulated absorption spectra of the cyclic aromatic structures in dependence on heptamer unit number n. For simplicity, the simulations were performed for the case of air ambient. Spectra are shifted upward for clarity. The red-dashed line highlights the Fano resonances of different structures. A clear resonance blue shift is visible with increasing n. The charge density plot of the Fano resonance for n = 6 is presented. The magnetic ring modes (highlighted using the blue bars) only occur when n is even. The charge density plots of the magnetic ring modes are displayed for n = 2, 4, and 6. The charge density plot of the complex Fano resonance mode at 1200 nm for n = 3 is also presented. Adapted from ref. [186].

Magnetic effects induced by the hybridization of gap plasmons was demonstrated in more complex MIM-like meta-molecules displaying a strong anisotropic optical response, such as nanostructured hyperbolic metamaterials [177]. Hyperbolic meta-molecules were also proved to enable an arbitrary control of scattering and absorption of light through the excitation of magnetic resonances, which offer a superior sensitivity and field enhancement, compared with the more conventional electric resonances [178]. It is worth mentioning that optical magnetism is an active research field, since magnetic resonances have been proved to provide a new degree of freedom for nanostructured systems, which can trigger unconventional nanophotonic processes, such as nonlinear effects [179] or electromagnetic field localization for enhanced spectroscopy [180] and optical trapping [181,182] (for more details, see the reviews by Calandrini et al. [183] and by Urzhumov and Shvets [184]). Important efforts have been spent in the development of metamaterials due to the difficulty to induce strong magnetic effects at optical frequencies by using natural materials or conventional plasmonic architectures. Although it is not the focus of this work to give a complete overview of optical magnetism at the nanoscale, we want to highlight some cases of particular interest in the framework of collective effects and coupling phenomena treated here. To achieve a larger control of the magnetic response, Verre et al. [185] proposed a very interesting approach to achieving optical magnetism in the visible. Being the magnetic resonances typically spectrally narrow, their applicability in metamaterial designs is quite limited (see Fig. 5). They show optical magnetism is rendered quasi-broadband through hybridization of the in-plane modes in plasmonic pentamers constructed from MIM units acting as interacting magnetic meta-atoms. In another interesting work, Liu et al. [186] studied heptamer structures [187,188] arranged in a waveguide geometry where antiphase magnetic plasmons are excited in adjacent fused heptamer units, giving rise to plasmonic anti-ferromagnetic behavior (Fig. 5). Shafiei et al. [189] demonstrated that a subwavelength plasmonic metamolecule consisting of four closely spaced gold nanoparticles supports strong magnetic response. Toma et al. demonstrated that coil-like dark modes and high-order plasmonic modes hybridization can generate strong magnetic effects in the mid-IR by using either planar asymmetric-disks or moon-integrated trimers [190,191] (see also Fig. 5). Another interesting approach to enhance magnetic effects, this time by using SLRs, was proposed by Tang et al. [192]. They used pairs of magnetic rods supporting displacement currents [193] as the elements in two-dimensional periodic arrays, showing a new way to enhance the magnetic fields at optical frequencies by using diffractive coupling of magnetic LSPRs. The authors predicted narrow-band hybrid plasmon/magnetic modes that occur because of the strong interaction between magnetic resonances and the collective surface lattice resonances of the array. Magnetic fields in the pairs of metal rods were 5-fold greater than the relevant fields of isolated individual pairs of metal rods. Morover, similarly to the approach used in ref. [145], Garcia-Camara et al. [194] have proposed hybrid arrays of electric and magnetic meta-atoms which exhibit simultaneously negative values for the electric permittivity ($\varepsilon$) and the magnetic permeability ($\mu$). When $\varepsilon<0$ and $\mu<0$, the phase of the wave moves in the opposite direction from the energy flow [195] and specific resonances can be excited in these systems (dipolar and quadrupolar resonances, both electric and magnetic), leading to a strong enhancement of the scattering efficiency. In particular, metamaterials with permittivity and permeability such that the simultaneous excitation of electric and magnetic dipolar resonances is allowed, can produce dramatic values for the scattered intensity. This approach, that is the combination of magnetic and electric LSPRs and their coupling leading to SLRs, is therefore useful to redirect scattered light, and might have a crucial impact on the development of photonic-based nanocircuitry [196].

**Conclusions and perspectives**

We have reviewed past and recent works on collective effects and coupling phenomena in resonant meta-atoms supporting both plasmonic and (opto-)magnetic properties. We have given an overview on their fundamental properties and applications, as well as on the present research challenges, such as the possibility to reach a real-time dynamic full-control of both magnetism (spins) and optics (light) at the same time. Although many interesting and inspiring studies have been done so far to bridge the gap, creating a strong and controllable interaction between light and magnetic fields is still deemed far from reaching the final solution. It is clear that harnessing light and spins through plasmons at the nanoscale may eventually lead to more effective ways to process and store information with photons and create different types of steerable optical elements with the outstanding ability typical of the human eye to adapt to changes and fluctuations in light characteristics. Many paths have still to be explored, first of all the use of exotic or non-conventional plasmonic modes such as optical anapoles [197,198], cavity [199,200] or dark modes [201-203] to push magneto-optical effects in magnetic metamaterials to their limits on the nanoscale. Moreover, the works on magneto-plasmonic crystals reported so far are all at normal incidence, while it is known that excitation of SLRs using both the in-plane and out-of-plane polarizations of the incident light brings interesting perspectives, such as the generation of high quality SLRs [204]. In this framework, it has been recently demonstrated that one-dimensional plasmonic crystals, which support very narrow surface lattice resonances [205], give rise to exceptionally large responses in higher order diffraction magneto-optical modes [206] or unexpected enhancement of magneto-optical transverse Kerr effect [207]. Moreover, as discuss along the article, femto-magnetism, which has been developed within the field of magnetism, can bring novel and exciting perspective on the excitation and control of magnetic dynamics on time scales comparable to or even much shorter than those of spin-lattice, spin-orbit, and exchange interactions [208-210]. For instance, due to opto-magnetic phenomena an intense laser pulse with a duration of 100 fs acts on the spins similar to an equivalently short effective magnetic field pulse up to 1 T, selectively exciting different magnetic resonances, triggering magnetic phase transitions or switching spins on a sub-picosecond time-scale. By investigating (opto-)magnetic metamaterials in this context may open the path towards new scenarios for applications of ultrafast opto-magnetic phenomena in magnetic storage and information processing technology [211]. Indeed, recent developments of nanolithography and other experimental approaches have disclosed that the interaction of light with magnetic nanostructures is qualitatively different from the regime in which the size of the illuminated media is much larger than the wavelength. In particular, the subwavelength regime of light–spin interaction revealed the plasmon-mediated enhancement and emergence of new magneto-optical phenomena. Novel possibilities of optical control of magnetism on the nanometer scale via plasmonic resonances have emerged as well it was shown that femtosecond laser pulses [$^{212}$] are able to generate magnons with nanometer wavelengths [213]. Moreover, intriguing new magnetic phenomena can be triggered in the THz region of the light spectrum [214], and fundamental efforts have been recently emerged in this direction to design metamaterials working in that frequency region to manipulate locally magnetic properties of matter [215,216]. Indeed, one of the missing pieces of this intricated puzzle is a systematic exploration of the fundamental properties of magneto-plasmonic metamaterials between the visible/near-infrared and the THz regions. Nevertheless, an interesting approach has been recently proposed to manipulate spin currents using mid-IR plasmons in magnetic micro-antenna and hole-array metamaterial platforms made of $Ni_{81}Fe_{19}$/Au multilayers, where the magneto-refractive effect linked to the giant magnetoresistance (GMR) present in the system allows a modification of the electrical resistivity upon the application of a small magnetic field [217]. This research field is still quite unexplored and might represent the point of contact between optics and magnetism at which the community as looked at so far. Finally, interesting perspectives may arise by studying nanostructured (opto-)magnetic metamaterials composed by non-conventional materials, such as magnetic semiconductors, garnets, and rare earth alloys with ultra-strong spin-orbit coupling effects, as well as the use of high-magnetic-anisotropy materials, possibly enabling to reach areal densities of 100 Tb per square-inch by using grains with sizes of the order of 1 nm [218] based on the use of single photons as information careers opening excellent opportunities also on quantum-based nanophotonic technologies [219].

## Acknowledgments

The Author acknowledges Paolo Vavassori, Alexandre Dmitriev, Irina Zubritskaya, Ruggero Verre, Matteo Pancaldi, Kristof Lodewijks, Luca Bergamini, Mikolaj Schmidt, Nerea Zabala, Javier Aizpurua, Mikko Kataja, Sara Pourjamal, Sebastian van Dijken, Denis Garoli, Yuri Gorodetski and Paolo Biagioni for the fruitful and inspiring discussions and/or collaborations on the topics treated in this review.